\documentclass[fleqn,twoside]{article}
\usepackage{espcrc2}
\usepackage{graphicx}
\usepackage[figuresright]{rotating}

\title{Identification of polymorphs of pentacene
\vspace{1pc}}

\author{Christine C. Mattheus \addressmark[grun]\address{Solid State Chemistry
        Laboratory, Materials Science Centre,
        University of Groningen, Nijenborgh 4,
        9747 AG Groningen, the Netherlands.} ,
        Anne B. Dros \addressmark, Jakob Baas \addressmark,
        Gert T. Oostergetel \addressmark, Auke Meetsma
        \addressmark,\\
        Jan L. de Boer \addressmark, and
        Thomas T. M. Palstra \addressmark \thanks{Corresponding author.
        Tel.: +31-50-363-4440; fax: +31-50-363-4441. E-mail:
        T.T.M.Palstra@chem.rug.nl}}

\begin{document}

\begin{abstract} Pentacene crystallizes in a layered structure
with a herringbone arrangement within the layers. The electronic
properties depend strongly on the stacking of the molecules within
the layers \cite{haddon}. We have synthesized four different
polymorphs of pentacene, identified by their layer periodicity,
$d$({\it 001}): 14.1, 14.4, 15.0 and 15.4~\AA. Single crystals
commonly adopt the 14.1~\AA\ structure, whereas all four
polymorphs can be synthesized in thin film form, depending on
growth conditions. We have identified part of the unit cell
parameters of these polymorphs by X-ray and electron diffraction.
The 15.0 and 15.4~\AA\ polymorphs transform at elevated
temperature to the 14.1 and 14.4~\AA\ polymorphs, respectively.
Using SCLC measurements, we determined the mobility of the
14.1~\AA\ polymorph to be 0.2 cm$^2$/Vs at room temperature.\\

\noindent \it {Submitted to Synthetic Metals on 21 august 2002,
accepted for publication on 19 september 2002.}

\end{abstract}

%\begin{keyword}
% keywords here, in the form: keyword \sep keyword
%Pentacene \sep Thin film \sep Single crystal \sep Crystal
%structure \sep Structural transition \sep SCLC
% PACS codes here, in the form: \PACS code \sep code

%\end{keyword}

\maketitle

\section*{Introduction}

\begin{figure}[htb]
   \centering
   \includegraphics[bb= 10 565 240 835, width=65mm]{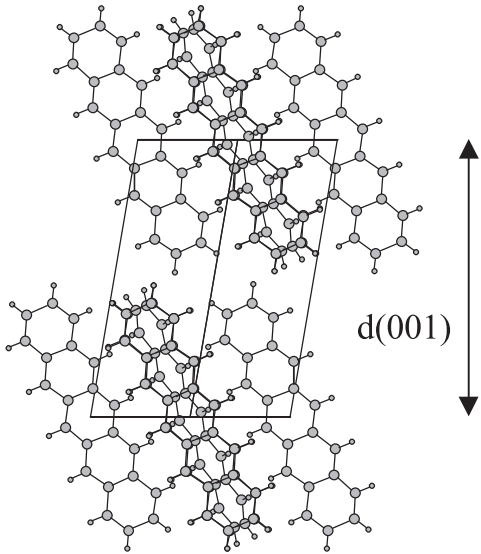}
   \caption{\emph{ The crystal structure of single crystalline
pentacene. The view along the $[{\it 1 \bar{1} 0}]$ axis shows the
layered structure, the unit cell is also indicated.}}
   \label{fig:penta}
\end{figure}

Pentacene has recently gained interest as a molecular conductor
with a very high electronic mobility. Mobilities of 1~cm$^2$/Vs
were reported at 300K, increasing to 10$^4$~cm$^2$/Vs at low
temperatures \cite{Bathole}. Moreover, evidence for
superconductivity up to 2~K in field effect devices was reported
\cite{Batsuper}. We study the origin of this high mobility and its
relation with the crystal structure, {\it i.e.} the stacking of
the molecules. The crystal structure and growth of pentacene has
been subject of a number of studies
\cite{Igor,Alex,Gundlach,Mina3,Holmes}. Pentacene crystallizes in
a layered structure with a herringbone arrangement within the
layers, see Fig.~\ref{fig:penta}. The layer periodicity, $d$(001),
is by far the largest repeat unit, and thus characteristic of the
structure. In the early sixties, Campbell {\it et al.} determined
the single crystal structure using a film method \cite{Camp61}.
These results were slightly modified \cite{Camp62}, but yielded
both a characteristic $d$({\it 001}) value of 14.4~\AA. In 1991
Minakata {\it et al.} reported on the growth and crystal structure
of thin films \cite{Mina1,Mina4}. They reported for thin films
$d$({\it 001})-spacings of 15.0 and 15.4~\AA. It was only noticed
by Dimitrakopoulos {\it et al.} \cite{Dimitra} that these thin
film polymorphs must be different from the single crystal. They
also observed for the first time the coexistence of two phases, of
14.4 and 15.4~\AA. The growth of these polymorphs was studied by
Schoonveld {\it et al.} \cite{Igor,Alex}. These phases are
generally referred to as the "single crystal" and "thin film"
phase. We argue the former name to be incorrect as more recent
redeterminations of the single crystal structure by Holmes {\it et
al.} \cite{Holmes}, Siegrist {\it et al.} \cite{Batcomm} and by us
\cite{ActaPolyPen} show the single crystal $d$({\it 001}) value to
be 14.1~\AA. Apparently there exist several polymorphs of
pentacene. Identification of these pentacene polymorphs is
especially important when the electronic properties are related to
the crystal structure. Recent band structure calculations show
that the electronic properties depend strongly on the particular
stacking of the molecules within the layers \cite{haddon}. In this
paper we discuss how the various polymorphs can be made, and what
their crystal structures are. We identify four different
polymorphs and determine by X-ray and electron diffraction their
unit cell parameters. Of one of the polymorphs we determine the
hole mobility, {\it via} space-charge-limited-current measurements
on a single crystal.

\section*{Experimental procedures}
Thin films of pentacene were prepared by evaporation in a high
vacuum environment of 10$^{-7}$~mbar. The source material, pure
pentacene was obtained from Aldrich and not further purified. It
was placed in a tantalum crucible and heated to 540~K. As
substrate thermally oxidized silicon ($a$-SiO$_{2}$) or kapton
(polyimide) was used. The substrate temperature was measured with
an Al-Cr thermocouple and the substrate could be heated or cooled.
The evaporation rate of the sublimed material was monitored by a
quartz oscillator and the layer thickness was determined using a
dektak. The evaporation was kept constant at a low rate of
0.1~nm/sec to ensure crystallinity. A shutter allowed the
evaporation rate to become stable before the substrate was exposed
to the pentacene flow.

Single crystals of pentacene were grown using a vapour transport
method \cite{Laudise}. A pyrex tube was thoroughly cleaned by
heating it under a stream of pure nitrogen gas. 200 - 400~mg of
unprocessed pentacene, obtained from Aldrich, was placed at one
end of the tube in a platinum crucible. The growth was performed
either under a stream of nitrogen gas mixed with hydrogen gas,
with a volume percentage of 5.1(1)\% hydrogen, or under an argon
flow. Transport gases were obtained from AGA with 5N purity for
N$_2$ and Ar and 4N5 purity for H$_2$. Great care was taken to
avoid contamination with O$_2$ and H$_2$O. Nitrogen and argon
gases were further purified over activated copper and alumina
columns. A temperature gradient was applied by resistively heating
two heater coils wrapped around the tube. Crystallization took
place some 300~mm from the sublimation point at a temperature of
approximately 490~K. Single crystals were also grown from a
solution of trichlorobenzene (TCB). Violet crystals were obtained
by slowly (four weeks) evaporating the TCB at 450~K, under a
stream of ultra pure nitrogen gas.

The X-ray diffraction patterns of the obtained single crystals
were measured with an Enraf-Nonius CAD-4 or a Bruker SMART APEX
diffractometer. Both diffractometers use monochromated Mo -
K$\alpha$ radiation. The structures are solved using SHELXS and
SHELXL \cite{Sheldrick}. The X-ray diffraction patterns of
pentacene thin films were measured using a D8 powder
diffractometer of Bruker A.G. in a Bragg-Brentano geometry, with
monochromatic Cu-K$\alpha_{1}$ radiation. A variable temperature
stage is used for in situ temperature dependent measurements
between 100~K and 650~K, in a vacuum of 10$^{-2}$~mbar. The
obtained spectra were analyzed using the TOPAS software package
\cite{Topas}. Electron diffraction (ED) measurements were
performed on a Philips CM200/FEG (200~kV) transmission electron
microscope. During the measurements the electron dose was held
low, to prevent radiation damage of the samples. Experiments were
performed in the Electron Microscopy group of the department of
biophysical chemistry in Groningen and in the National Centre for
HREM in Delft.

\section*{Growth of the different polymorphs on thin films}
\begin{figure}
   \centering
   \includegraphics[bb= 0 650 235 840, width=65mm]{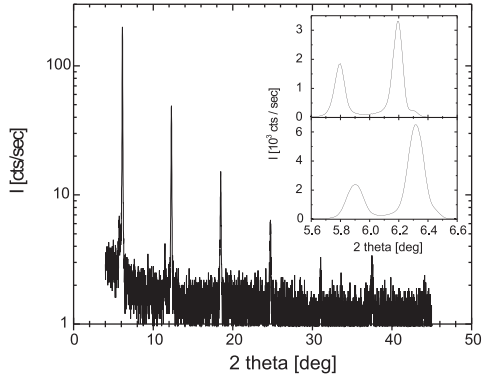}
   \caption{\emph{X-ray diffraction pattern for a pentacene
thin film grown at 353~K on thermally oxidized $a$-SiO$_{2}$. The
reflections can be indexed as (00l), with a $d$({\it 001}) spacing
of 14.42~\AA.  Inset, top: part of a X-ray spectrum of a thin film
grown on $a$-SiO$_{2}$. The observed reflections correspond to
$d$(001) values of 15.4 and 14.4~\AA. This specific (2.5~$\mu$m
thick) sample also contains a small amount of the 14.1~\AA\ phase,
as calculated from the shoulder of the reflection at
6.19\,$^{\circ}$. Inset, bottom: part of a X-ray spectrum of a
thin film grown on kapton. The reflections correspond to $d$({\it
001}) values of 15.0 and 14.1~\AA. The broader peak shape for the
films grown on kapton indicates a larger mosaic spread of these
films.}}
   \label{fig:Xraypoly}
\end{figure}

In the literature different $d$({\it 00l}) values are reported for
pentacene thin films \cite{Igor,Mina1,Mina4,Laquin}. To
investigate this, we have grown thin films of pentacene, using
different growth conditions. A typical X-ray diffraction pattern
of a pentacene thin film is shown in Fig.~\ref{fig:Xraypoly}. The
peaks can be indexed as ${\it 00l}$ reflections, indicating a
strong preferential alignment with the $[{\it 00l}]$~vector
perpendicular to the substrate.

In the inset of Fig.~\ref{fig:Xraypoly}, parts of two diffraction
patterns of two thin films, grown in different ways, are shown. We
observe reflections centered at 5.80, 5.92, 6.19 and 6.33 degrees
2$\theta$. Analysis of the entire spectra shows the reflections to
correspond to $d$({\it 001})~spacings of 14.1, 14.4, 15.0 and 15.4
\AA. This result indicates that at least four polymorphs of
pentacene are present in the various thin films.

Typically, two polymorphs are observed on one thin film sample, as
shown in the inset of Fig.~\ref{fig:Xraypoly}. However, single
phased thin films can be grown. In literature the growth of the
14.4 and 15.4~\AA\ polymorphs was reported to depend on the film
thickness and the substrate temperature~\cite{Igor,Alex}. We
observe that the growth of the polymorphs depends also on the
nature of the substrate. The 14.4 and 15.4~\AA\ polymorphs grow on
$a$-SiO$_{2}$ substrates, whereas the 14.1 and 15.0~\AA\
polymorphs grow on kapton. The top of the inset in
Fig.~\ref{fig:Xraypoly} shows a part of a spectrum of a film grown
on  $a$-SiO$_{2}$. In this spectrum we observe the 14.4 and
15.4~\AA\ polymorphs. At the bottom of this figure, the pattern of
a film grown on kapton is shown. Here, peaks correspond to the
14.1 and 15.0~\AA\ polymorphs.

The 15.4~\AA\ phase can be grown single phased if the film
thickness remains below a certain critical value, which depends on
the substrate temperature. This polymorph is typically grown at
room temperature. The 14.4~\AA\ phase is observed above this
critical thickness, and at an elevated substrate temperature. The
fraction of this polymorph increases with the film thickness
\cite{Igor,Alex}. We observe that this dependence on the substrate
temperature and the film thickness also holds for the 15.0 and
14.1~\AA\ polymorphs. Here, the 15.0~\AA\ polymorph is stable for
small film thicknesses and low substrate temperatures.

Typical growth conditions to obtain single-phased films are listed
in Table~\ref{tbl:thinfilms}. We have grown single phased films of
the 15.0 and 15.4~\AA\ phases at ambient temperature with a film
thickness of approximately 50~nm, on a kapton and $a$-SiO$_{2}$
substrate, respectively. The 14.1 and 14.4~\AA\ phases were
observed single-phased on films grown at 370~K, with a thickness
of $\sim$150~nm, on substrates of kapton and $a$-SiO$_{2}$,
respectively.

\begin{table}
 \centering
 \begin{tabular}{l|l|l}
                   &  $50$ nm  & $150$ nm   \\
                   &   at $300$ K               &   at $350$ K                     \\\hline
   $a$-SiO$_{2}$   & 15.4             & 14.4                    \\\hline
   kapton          & 15.0             & 14.1                    \\
 \end{tabular}
 \caption{\emph{Typical conditions for single-phase growth of the specific pentacene polymorphs.}}
 \label{tbl:thinfilms}
\end{table}

The spectrum on the top of the inset in Fig.~\ref{fig:Xraypoly}
was of a 2 $\mu$m thick film. In this spectrum we observe not only
two, but three polymorphs of pentacene. Besides the 15.4 and
14.4~\AA\ polymorphs, the shoulder of the peak at 6.19$^\circ$
2$\theta$ could be recognized as belonging to the 14.1~\AA\
structure. This suggests that the growth of the different phases
on an $a$-SiO$_{2}$ substrate is as follows. In the first $\sim$30
monolayers the 15.4~\AA\ polymorph is formed, if the film becomes
thicker also the 14.4~\AA\ structure appears. If the film becomes
extremely thick ($>$1300 monolayers), also the 14.1~\AA\ phase is
observed. The 14.1~\AA\ structure is observed in single crystals
and on thin films grown on polyimide. This substrate does,
therefore, not seem to influence the growth. However, the
appearance of the 14.1~\AA\ polymorph on thick films on a
$a$-SiO$_{2}$ substrate indicates that the 14.4 and 15.4~\AA\
polymorphs are substrate induced.

\section*{Temperature dependence}

\begin{figure}
   \centering
   \includegraphics[bb= 0 675 235 840, width=65mm]{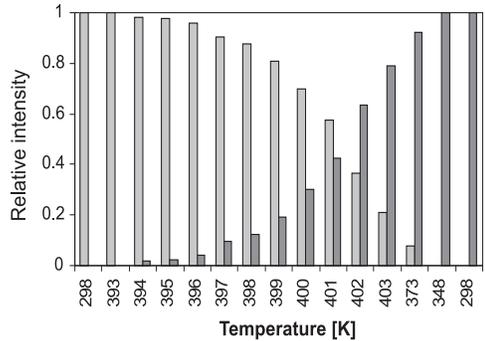}
   \caption{\emph{The relative intensity of the 15.4 and
14.4~\AA\ phases for a temperature sequence. Light grey: 15.4~\AA,
dark grey: 14.4~\AA.}}
   \label{fig:htxrd155}
\end{figure}

We studied the possibility of structural phase transitions between
143 and 423~K. Firstly, a film of the 15.4~\AA\ structure was
measured at ambient temperature, slowly heated to 403~K and cooled
down again. An irreversible change of the 15.4~\AA\ phase into the
14.4~\AA\ phase was observed. This structural change as function
of temperature is shown in Fig.~\ref{fig:htxrd155}. In this figure
relative intensities are shown, because above 370~K sublimation of
the material started. This decreases the measured intensity and
makes the structural change difficult to observe.

Heating a film containing both the 15.0 and the 14.1~\AA\ phases,
to 423~K, showed behaviour analogous to the 15.4~\AA\ phase. At
elevated temperatures the 15.0~\AA\ phase changed irreversibly
into the 14.1~\AA\ structure.

We cooled and heated the 14.1~\AA\ polymorph in single and
poly-crystalline form. A full single crystal data set was measured
at 90~K. However, no phase transition was observed. From the
temperature dependence of X-ray spectra of powder of the 14.1~\AA\
polymorph, in the range of 143~K to 475~K, we derive anisotropic
linear coefficients of thermal expansion: $\alpha_a$ =
-19~$\cdot$~10$^{-6}$~K$^{-1}$, $\alpha_b$ =
64~$\cdot$~10$^{-6}$~K$^{-1}$, and $\alpha_c$ =
53~$\cdot$~10$^{-6}$~K$^{-1}$.

A film containing only the 14.4~\AA\ polymorph was slowly heated
to 475~K, but no structural transition was observed. We cooled a
sample containing both the 15.4 and the 14.4~\AA\ polymorphs to
143~K. No structural changes were observed during the cooling
process. Also a film containing the 15.0~\AA\ and 14.1~\AA\
polymorphs did not show any structural transition during cooling.
From these experiments it can be concluded that the 14.1 and the
14.4~\AA\ phases are thermally the most stable ones. The 15.0 and
15.4~\AA\ structures can, at elevated temperatures, be transformed
into the 14.1 and 14.4~\AA\ phases, respectively.

It was already observed by Gundlach {\it et al.} that the
15.4~\AA\ phase can be transformed into the 14.4~\AA\ polymorph
\cite{Gundlach}. They exposed their films of the 15.4~\AA\ phase
to solvents like acetone, isopropanol, and ethanol (in which
pentacene does not dissolve) and noticed that the 15.4~\AA\ phase
almost completely changed to the 14.4~\AA\ phase. We performed the
same experiments and observed that indeed the 15.4~\AA\ phase
disappeared from the spectra. Since we lost some material in the
process, it was not clear whether it was transformed into the
14.4~\AA\ phase or had become amorphous. The same observation was
made for the 15.0~\AA\ polymorph, which transformed after exposing
the film for $\sim$3 minutes to ethanol.

\section*{The 14.1~\AA\ polymorph: single crystals}

Besides the growth of thin films, also single crystals were grown.
Single crystals, grown in various ways, yield exclusively the
14.1~\AA\ polymorph. Crystal growth from vapour transport yields
almost centimeter sized violet crystals, in different forms:
platelets and needles. These pentacene single crystals were
measured on the CAD-4 diffractometer. Needle shaped crystals were
observed to grow along the [{\it 1$\bar{1}$0}] axis, whereas the
platelets grow in the $ab$ plane. Both had a $d$({\it 001}) value
of 14.1~\AA. Pentacene single crystals grown from a solution of
trichlorobenzene were analyzed on the APEX diffractometer. This
yielded the same crystal structure as for the crystals grown using
the vapour transport technique. The unit cell parameters are given
in Table~\ref{tbl:datarealspace} and more details are reported
elsewhere \cite{ActaPolyPen,Thesis}.

The vapour transport experiments did not result in single crystals
of pentacene exclusively. The employed transport gases, and their
flow rate, were of significant influence on the crystals. At a
slightly lower temperature (480~K) than where the violet pentacene
crystals crystallized, also red needles were grown. If excess
hydrogen gas was used, more red crystals were observed. The
crystallization point was $\sim$~50~mm apart from the centre of
the pentacene growth, which allowed physical separation of the two
types of crystals. Single crystal diffractometry showed it to be a
hydrogenated form of pentacene: 6,13-dihydropentacene, with two
methylene groups at opposite sides of the middle ring. The
structure can be described as a monoclinic unit cell containing
four non-planar hydrogenated pentacene molecules and two planar
pentacene molecules. The detailed crystal structure is described
elsewhere \cite{Thesis}.

At lower hydrogen content, or if no ultra pure inert transport gas
was used, yellow-brown coloured crystals could be observed. They
condense at a slightly higher temperature than the pure pentacene
crystals, at $\sim$~520~K. Analysis on the APEX diffractometer
proved it to be an oxidized form of pentacene,
6,13-pentacenequinone, with two carbonyl groups at opposite sides
of the middle ring. The unit cell is monoclinic, and described
elsewhere \cite{Thesis}, in agreement with literature
\cite{Dzyab}.

\section*{The 14.4~\AA\ polymorph: powder diffraction}

\begin{figure}
   \centering
   \includegraphics[bb= 0 685 235 840, width=65mm]{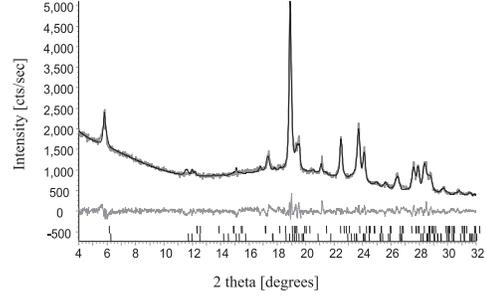}
   \caption{\emph{Observed (grey) and simulated (black)
diffraction pattern of a 2.5~$\mu$m thick film of pentacene, grown
on $a$-SiO$_{2}$ at 363~K, and removed from its substrate. The
pattern consists of a majority (91\%) 14.4~\AA\ polymorph and a
minority (9\%) 14.1~\AA\ polymorph. The difference spectrum is
shown by the gray line under the spectrum. All reflections could
be indexed, but the parameter set was to large to fit the
intensities and obtain fractional coordinates for the 14.4~\AA\
structure.}}
   \label{fig:powd145}
\end{figure}

The 14.4~\AA\ polymorph was grown by vacuum sublimation on
$a$-SiO$_{2}$ at 363~K. The Bragg-Brentano diffraction pattern
indicated that the 2$\mu$m thick film contained mostly the
14.4~\AA\ structure, with small amounts of 15.4 and 14.1~\AA\
polymorphs. The 15.4~\AA\ phase was removed using Gundlachs
method~\cite{Gundlach} of dipping the film in ethanol. A new
diffraction pattern confirmed this. The film was mechanically
removed from the substrate and ground into a powder. The powder
was dispersed on Scotch tape and remeasured in a transmission
geometry on the Bruker D8 diffractometer. The powder diffraction
pattern at room temperature is shown in Fig.~\ref{fig:powd145}.

Clearly, reflections other than ${\it 00l}$ can be observed. The
unit cell parameters were analyzed, while the fractional
coordinates were not further refined. The pattern can be fitted
with a mixture of two phases. The majority phase (91\%) was found
to have unit cell parameters which are listed in
Table~\ref{tbl:datarealspace}, and yields a $d$({\it 001}) of
14.37~\AA. A minority phase could be indexed with the 14.1~\AA\
single crystal unit cell.

\begin{sidewaystable}
%\begin{table}
\centering
\begin{tabular}{c|cccccc|c}
              & a        & b        & c         & $\alpha$  & $\beta$  & $\gamma$   & $d$(001)\\\hline
  Single Crystal 300K  & 6.266(1) & 7.775(1) & 14.530(1) & 76.475(4) & 87.682(4) & 84.684(4) & 14.12 \\
  Single Crystal 90K & 6.239(1) & 7.636(1) & 14.330(2) & 76.978(3) & 88.136(3) & 84.415(3) & 13.96 \\
  Powder diff.& 6.485(1) & 7.407(2) & 14.745(4) & 77.25(2)  & 85.72(2)  & 80.92(2)  & 14.37 \\
\end{tabular}\\
  \caption{\emph{The lattice parameters of the 14.1 and 14.4 \AA\ pentacene polymorphs.
  The data of the single crystal polymorph are listed for both 90~K and 300~K.}}
  \label{tbl:datarealspace}
%\end{table}
\end{sidewaystable}

\section*{The 15.0 and 15.4~\AA\ polymorphs: electron diffraction}

Because the 15.0 and 15.4~\AA\ polymorphs can only be grown for
small film thickness, insufficient sample was available for X-ray
powder diffraction. Here, we used electron diffraction (ED) to
obtain structural information.

We performed electron diffraction experiments in a transmission
geometry. ED has the advantage above X-ray based techniques, that
the incident beam can be focussed. The scattering cross section is
large, and can thus probe very small areas. In our ED experiments
an electron beam with a diameter of 0.3~$\mu$m is used. Atomic
force microscopy pictures of thin films show that films made at
room temperature typically have a crystal size of 0.2 -
0.3~$\mu$m. Films grown at a higher temperature are more
crystalline, and exhibit crystal sizes up to 1~$\mu$m. Electron
diffraction offers thus the possibility to perform diffraction
experiments on one single crystal area instead of a
polycrystalline sample.

\begin{figure}
   \centering
   \includegraphics[bb= 0 565 240 840, width=65mm]{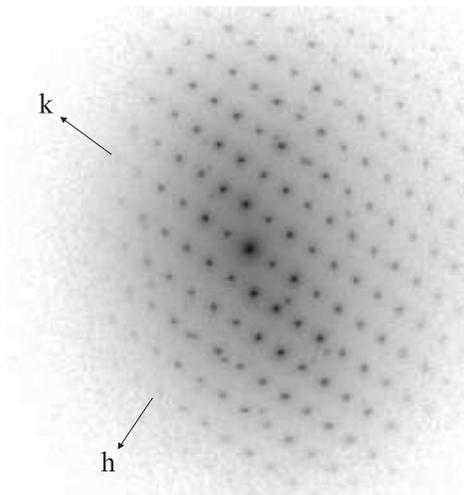}
   \caption{\emph{Transmission electron microscopy pattern. A
typical projection of a $({\it hk0})$ plane is shown.}}
   \label{fig:tem780}
\end{figure}

Samples for ED were prepared by evaporating a pentacene layer on a
copper grid, which is covered with a carbon film. The pentacene
layer of $\sim$100~nm, was evaporated on a substrate which was
held at 345~K. When a single crystal area is located on the
sample, the ED image exhibits spots instead of circles, as shown
in Fig.~\ref{fig:tem780}. X-ray diffraction experiments showed
strong preferential orientation with the [{\it 00l}] axis.
Therefore it is likely that the spot patterns observed with ED, in
transmission, are generated by the ({\it hk0})~planes. From this
diffraction pattern, the reciprocal lattice parameters $a^{*}$,
$b^{*}$ and $\gamma^{*}$ can be determined. This measurement was
calibrated by measuring a gold sample as reference. Various
diffraction measurements, on different locations on the same
sample, resulted in three different sets of ${\bf a^{*}}$, ${\bf
b^{*}}$ and $\gamma^{*}$ values. We observed on an edge of the
same sample also one ${\bf c^{*}}$ value. The results are listed
in Table~\ref{tbl:struc}. The reciprocal lattice data of the
single crystal and the ED results calculated from Minakata {\em et
al.} \cite{Mina3} have been added to this table. Note that the
difference between ${\bf a^*}$ and -${\bf a^*}$ or ${\bf b^*}$ and
-${\bf b^*}$ could not be observed. Therefore a $\gamma^*$ value
of $89^{\circ}$ can also be 91$^{\circ}$.

\begin{sidewaystable}
%\begin{table}
\centering
\begin{tabular}{c|cccccc|c}
 & $a^{*}$ & $b^{*}$ & $c^{*}$  & $\alpha^{*}$ & $\beta^{*}$ & $\gamma^{*}$ & $d$(001)\\\hline
  SXD  & 0.1603 & 0.1328 & 0.0708 & 103.374   & 91.1114  & 94.91     & 14.12 \\
  ED1  & 0.1610 & 0.1319 &        &           &          & 89.5      &       \\
  XRD  & 0.1563 & 0.1399 & 0.0696 & 102.25    & 92.37    & 98.4      & 14.37 \\
  ED2  &        &        & 0.0694 &           &          &           & 14.4  \\
  ED3  & 0.173  & 0.134  &        &           &          & 89        &       \\
  Minakata$^{(1)}$ & 0.174  & 0.135 &    &           &          & 90        &       \\
  ED4  & 0.180  & 0.140  &        &           &          & 89.5      &       \\
  XRD  &        &        &        &           &          &           & 15.0 \\
  XRD  &        &        &        &           &          &           & 15.4 \\
\end{tabular}\\
 \caption{\emph{The reciprocal lattice parameters of pentacene, values are
 in \AA$^{-1}$. $d$({\it 001}) values are in \AA. Each line lists an experiment,
 performed on different samples or different parts of a sample. SXD is single
 crystal X-ray diffraction, XRD is powder X-ray diffraction, ED means electron
 diffraction. (1):~data calculated from Minakata {\em et al.}
\cite{Mina3}.}} \label{tbl:struc}
%\end{table}
\end{sidewaystable}

We observe all four polymorphs of pentacene on the same sample. In
the first set of observed values (ED1, see Table~\ref{tbl:struc})
the ${\bf a^*}$ and ${\bf b^*}$ values correspond to the single
crystal data. The $\gamma^*$ value, however, deviates beyond the
measuring precision from the value determined with single crystal
diffraction. We have no explanation for this difference and
neglect it. The 14.4~\AA\ phase was present, as an observed ${\bf
c^*}$ value (ED2) indicates the presence of this polymorph. The
third set of parameters observed in our ED experiments (ED3) is in
good agreement with the values observed by Minakata {\it et al}.
However, their reported ${\bf c^*}$ value was measured on a sample
grown under the same conditions, but on a different substrate.
Since we observed the nature of the substrate to be of crucial
importance to the growth of a specific polymorph, we omit their
reported value of ${\bf c^*}$. ED4 in Table \ref{tbl:struc} lists
a third set of ${\bf a^*}$, ${\bf b^*}$ and $\gamma^*$ values.
These sets of data, ED3 and ED4, will belong to the 15.0 and
15.4~\AA\ polymorphs.

In these experiments the electron beam was perpendicular to the
substrate. By tilting the sample other reciprocal lattice vectors
can be brought in diffraction condition. If sufficient {\it hkl}
spots are identified, the unit cell can be constructed. However,
the tilting may not change the small single crystalline area under
study. Presently, tilting did affect the sample position, and we
could not ascertain that the diffraction pattern belonged to the
same polymorph. Furthermore, after a short exposure to the
electron beam, the material suffered from radiation damage. Future
modifications of the apparatus will allow to tilt the sample while
keeping the same crystal in the beam.

\section*{Charge transport properties of the 14.1~\AA\ polymorph.}

We have used space-charge-limited-current measurements on single
crystals of pentacene to determine the hole mobility of the
14.1~\AA\ polymorph. Single crystals were grown under an argon
atmosphere as described earlier. Rectangular shaped crystals
(platelets) were glued onto glass substrates using GE-varnish.
Contacts were applied using a shadow mask. Stripes of 10~nm
titanium and 40~nm gold were evaporated in a high vacuum
environment. Distances between the stripes were 50 or 75~$\mu$m.
Measurements were performed in the $ab$ plane. Platinum wires were
connected to the stripes with silver epoxy. The epoxy was cured at
70$^\circ$C in air for 10 minutes. The crystals were exposed to
air for approximately one hour before they were placed in vacuum.
Samples were measured in darkness and a vacuum of
2~$\cdot$~10$^{-7}$ mbar. As measuring device a Keithley 2410 high
voltage sourcemeter was used. The $J$($E$) characteristics were
measured using a pulsed staircase sweep with a pulsewidth of
0.3~sec.

\begin{figure}
   \centering
   \includegraphics[bb= 15 625 255 820, width=65mm]{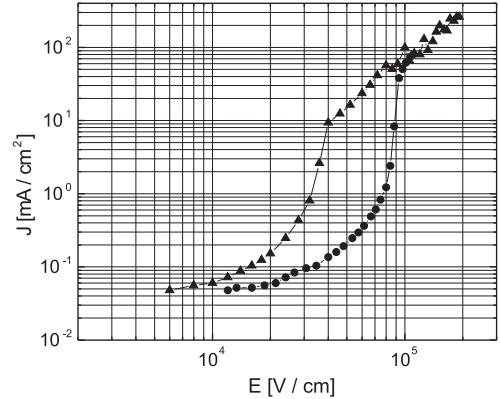}
   \caption{\emph{Current-voltage characteristic of single
crystalline pentacene (14.1~\AA\ polymorph). $\triangle$: gap of
50~$\mu$m, $\bullet$ : gap 75~$\mu$m.}}
   \label{fig:sclc}
\end{figure}

The J(E) measurements can seen in Fig.~\ref{fig:sclc}. The results
of these current-voltage measurements can be described using the
standard semiconductor band model~\cite{kao}. We assume holes to
be the majority carriers in the crystals. From analysis of the
curves \cite{kao,Bathole} we calculated the hole mobility, trap
density, the energy of the trap level and the acceptor density:
$\mu$ = 0.2 cm$^{2}$/Vs, N$_{t}$ = 2 $\cdot$ 10$^{13}$ cm$^{-3}$,
E$_{t}$ = 580 meV and N$_{a}$ 4 $\cdot$ 10$^{12}$ cm$^{-3}$.

\section*{Conclusion}

Pentacene has been shown to crystallize in four different
polymorphs. These different structures can be identified by their
$d$({\it 001}) value, which is characteristic for the structure.
Values of 14.1, 14.4, 15.0 and 15.4~\AA\ have been observed.
Single crystals commonly adopt the shortest $d$({\it 001})~spacing
of 14.1~\AA. All four polymorphs can be made as thin films. The
growth of the different polymorphs as thin films was shown to
depend strongly on the substrate, substrate temperature and film
thickness. The 14.4 and 15.4~\AA\ polymorphs appear to be induced
by the substrate.

Temperature dependent X-ray diffraction measurements showed that
the 14.1 and 14.4~\AA\ phases are the most stable. The 15.0 and
15.4~\AA\ structures are less stable and can be changed into 14.1
and 14.4~\AA, respectively, at elevated temperatures or by
exposing them to solvents. There is no indication of a structural
phase transition in the 14.1 and 14.4~\AA\ phases in the
temperature range of 143 - 473~K.

Pentacene single crystals have been grown, both by vapour
transport methods and from solution. The crystal structure that we
report \cite{ActaPolyPen} agrees with the analysis of Holmes {\it
et al.} \cite{Holmes} and Siegrist {\it et al.} \cite{Batcomm},
but is distinctly different from the data reported by Campbell
{\it et al.} \cite{Camp61,Camp62}.

Pentacene is during crystal growth sensitive to both oxidation and
hydrogenation, resulting in single crystals of dihydropentacene
and pentacenequinone. The methylene and carbonyl groups are
located at opposite sides of the middle ring, which seems the most
reactive.

Powder diffraction experiments were used to obtain the unit cell
parameters of the 14.4~\AA\ polymorph. Electron diffraction
experiments revealed for three polymorphs the ${\bf a^*}$, ${\bf
b^*}$ and $\gamma^*$ values.

Using SCLC measurements we determined the mobility of the
14.1~\AA\ polymorph of pentacene to be 0.2~cm$^{2}$/Vs.
Furthermore, these measurements show our single crystals to be
very pure, having only 10$^{13}$ traps cm$^{-3}$.

\section*{Acknowledgements}
We acknowledge stimulating discussions with G. Wiegers. We thank
H. Zandbergen for stimulating contributions for the electron
diffraction results.

\end{document}